\documentclass[reprint,amsmath,amssymb,aps,prl]{revtex4-1}

\usepackage{graphicx}
\usepackage{dcolumn}
\usepackage{bm}
\usepackage{color}
\usepackage{braket}  
\usepackage{subfigure}

\begin{document}

\title{Quantum $\mathcal{PT}$-Phase Diagram in a Non-Hermitian Photonic Structure}

\author{Xinchen Zhang$^{1}$}
\thanks{These two authors contribute equally.}
\author{Yun Ma$^{1}$}
\thanks{These two authors contribute equally.}
\author{Qi Liu$^{1,2}$}
\author{Nuo Wang$^{1}$}
\author{Yali Jia$^{1}$}
\author{Qi Zhang$^{1,6}$}
\author{Zhanqiang Bai$^{7}$}
\author{Junxiang Zhang$^{8}$}
\author{Qihuang Gong$^{1,2,3,4,5}$}
\author{Ying Gu$^{1,2,3,4,5}$}
\email{ygu@pku.edu.cn}

\affiliation{$^1$State Key Laboratory for Mesoscopic Physics, Department of Physics, Peking University, Beijing 100871, China\\
$^2$Frontiers Science Center for Nano-optoelectronics $\&$  Collaborative Innovation Center of Quantum Matter $\&$ Beijing Academy of Quantum Information Sciences, Peking University, Beijing 100871, China\\
$^3$Collaborative Innovation Center of Extreme Optics, Shanxi University, Taiyuan, Shanxi 030006, China\\
$^4$Peking University Yangtze Delta Institute of Optoelectronics, Nantong 226010, China\\
$^5$Hefei National Laboratory, Hefei 230088, China\\
$^6$Institute of Navigation and Control Technology, China North Industries Group Corporation, Beijing 100089, China\\
$^7$School of Mathematical Sciences, Soochow University, Suzhou, 215006, China\\
$^8$Zhejiang Province Key Laboratory of Quantum Technology and Device, Department of Physics, Zhejiang University, Hangzhou 310027, China}

\date{\today}
\begin{abstract}
Photonic structures have an inherent advantage to realize PT-phase transition through modulating the refractive index or gain-loss.
However, quantum PT properties of these photonic systems  have not been comprehensively studied yet.
Here, in a  bi-photonic structure with loss and gain simultaneously existing, we analytically obtained the quantum PT-phase diagram under the steady state condition.
To characterize the PT-symmetry or -broken phase, we define an Hermitian exchange operator  expressing the exchange between quadrature variables of two modes. 
If inputting several-photon Fock states into a PT-broken bi-waveguide splitting system, most photons will concentrate in the dominant waveguide with some state distributions. 
Quantum PT-phase diagram paves the way to the quantum state engineering, quantum interferences, and  logic operations in non-Hermitian photonic systems.
\end{abstract}

\maketitle

{\it{Introduction.}}
Open quantum system generally exchanges the energy with the external environment, i.e., it is non-Hermitian.
With varying some specific parameters in non-Hermitian parity-time (PT) system, there exist exceptional points (EPs) from PT-symmetry to broken, where the eigenvalues and corresponding eigenvectors simultaneously coalesce \cite{1998.PhysRevLett}. 
Various theoretical works related to PT-symmetry are proposed  \cite{TH1,TH2,TH3}, exhibiting some interesting phenomena, such as optical solitons and Bloch oscillations in periodical potentials  \cite{Period2, Period3}, edge-gain effect and gain-loss-induced skin modes in topological systems  \cite{Topo3, Topo5}.
Simultaneously, PT-symmetry and broken behaviors are experimentally realized in atomic and trapped ion systems  \cite{Atom, EXP5}, acoustic medium \cite{Acoustic3}, electronic circuit \cite{EXP1}, photonic lattice \cite{EXP2}, quantum optical systems \cite{EXP3, EXP4, EXP6}.

In addition to above mentioned systems, photonic structures are good candidate to realize PT-symmetry or broken through modulating the refractive index or gain-loss \cite{2007.OpticsLett, 2019.science}.
Owing to the similarity between the Schrodinger equation and the paraxial optical equation \cite{2017.NatPhoto}, photonic structures have an inherent advantage for realizing PT-symmetry. 
Both optical waveguides \cite{2009.PhysRevLett,2010.NatPhys} and whispering gallery microcavities \cite{2014.NatPhoto,2014.NatPhys} can construct PT-symmetric system by two-mode coupling with gain and loss. 
Besides, PT-symmetry has been observed in metasurface \cite{2014.PhysRevLett.(Manifestation} and periodically modulated refractive index material \cite{2011.PhysRevLett,2011.Science}. 
Because of the non-reciprocal property in PT-broken and enhanced sensitivity at EPs, PT-symmetric optics can be applied in optical isolation devices \cite{2014.NatPhoto}, sensing \cite{2014.PRL}, laser \cite{2014.science.(Loss-induced,2014.science.(Single}, and chiral optics \cite{2016.PNAS}.

However, previous studies on PT-symmetric photonic structures are almost limited to classical optics, where loss and gain in the same mode can cancel each other and be considered as an average effect.
While in quantum PT system, the role of loss and gain is different: the gain while generating photons will bring some noise, but the loss while annihilating photons can not lower any noise and even cause vacuum noise. 
These two irreversible processes inevitably produce different kinds of quantum jumping, leading to some interesting quantum behaviors. With the consideration of quantum jumping, people studied the saturation effects on the noise and entanglement \cite{2019.PRA.(Scully-Lamb),2019.PRA.(Effect)}, the positions and characteristic of EPs \cite{2019.PRA.(QuantumExceptional),2020.PRA.(Quantum)}, and the switching between PT and anti-PT systems \cite{2020.PRA.(Liouvillian)} in non-Hermitian gain-loss coupled cavities. Until now, there is no panoptic study on the PT-broken behavior of full gain-loss parameter space, i.e., quantum PT-phase diagram. Once this phase diagram is obtained, 
people can use photonic structures to engineer the quantum state and to realize the quantum logic operation, especially when PT-symmetry is broken.

\begin{figure*}[htb]
  \centering
  \includegraphics[width=\textwidth]{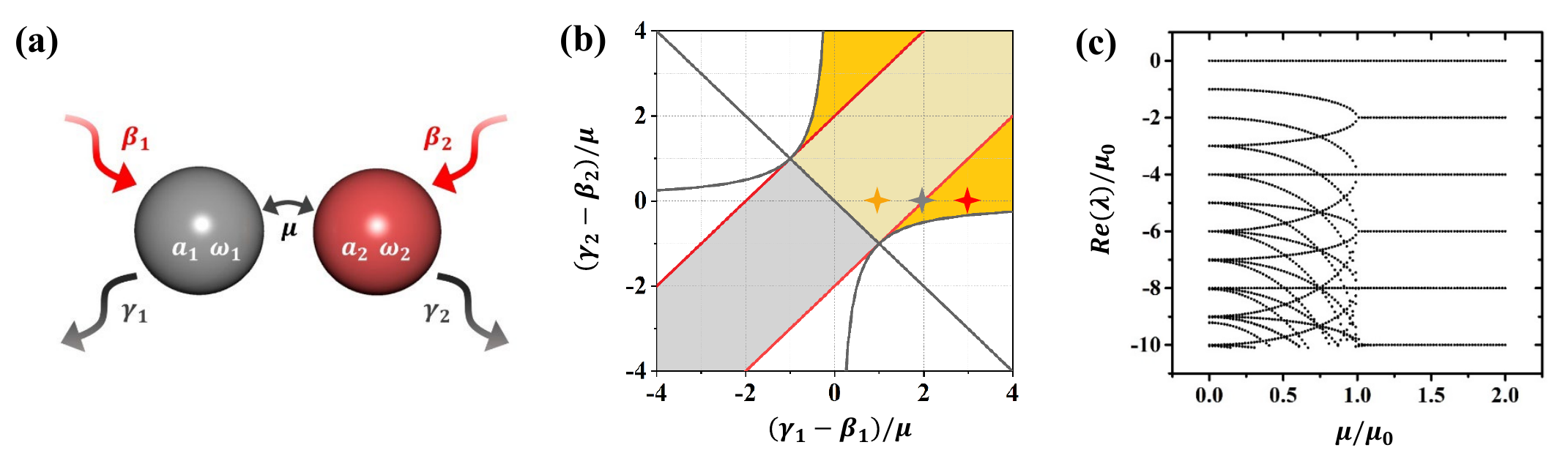}\\
  \caption{\label{fig:Fig1}
  PT-phase diagram of two quantum photonic cavities.
  (a) Schematic diagram of bi-photonic structures with  the coupling coefficient $\mu$ and the loss rate $\gamma_{j}$, gain rate $\beta_{j}$ for $j=1, 2$.  Here we let $\omega_{1}=\omega_{2}$.  
  (b) PT-phase diagram with the steady state regime. 
EPs satisfying $(\beta_{1}-\gamma_{1})/\mu-(\beta_{2}-\gamma_{2})/\mu=\pm2$ are shown as red lines. 
The regime between two EP lines is PT-symmetry while outside of EP lines is PT-broken. 
The yellow part is the regime in which the steady state exists.
Yellow star: PT-symmetry; Grey star: EP; Red star: PT-broken. 
(c) Real parts of eigenvalues of Liouvillian $\mathcal{L}(\mu)$ with the EP at $\mu=\mu_0 $. Here $\gamma_1=3.1\mu_0, \beta_1=0.1\mu_0, \gamma_2=1.1\mu_0$, and $\beta_2=0.1\mu_0$, respectively. 
}
\label{fig1}
\end{figure*}

 In this Letter, we analytically obtain the quantum phase diagram of PT-symmetry or broken in bi-photonic cavities with both gain and loss simultaneously existing [Fig. 1(a)]. 
For the consideration of reality, the steady state regime under the weak gain is identified.
To characterize the transition from PT-symmetry to broken, we define the exchange operator with exchanging the quadrature variables between two modes.
Then, based on PT-symmetry or broken regime in the above phase diagram, we explore the quantum splitting behaviors with the discrete variable of several photons.
If inputting Fock states into a PT-broken bi-waveguide splitting system, most photons concentrate in the dominant waveguide with some state distributions; while in the PT-symmetry situation, photons are alternately distributed in two waveguides with the variation of time.
The phase diagram with full parameter space will give us an in-depth understanding in quantum PT-symmetric system.
It is also the basis to study the quantum state fabrication, quantum interferences, and logic operations in non-Hermitian quantum photonic systems.
  
{\it{Quantum PT-phase diagram with steady state regime.}}
Consider bi-photonic cavities with loss and gain simultaneously existing [Fig. 1(a)],  when we let  $\omega_1=\omega_2=\omega$,
whose Hamiltonian is 
\begin{equation}
\hat{H}=\hbar \omega \hat{a}_{1}^{\dagger} \hat{a}_{1}+\hbar \omega \hat{a}_{2}^{\dagger} \hat{a}_2+\hbar \mu (\hat{a}_{1}^{\dagger} \hat{a}_{2}+\hat{a}_{2}^{\dagger}\hat{a}_{1})
\end{equation}
where $\hat{a}_j$ and $\hat{a}_j^{\dagger}$ ($j=1,2$) are the boson annihilation and creation operator, respectively, and $\mu$ is the coupling strength between two cavities. 
With the weak gain and weak incident light, the gain saturation effect can be neglected  \cite{1974.LaserPhysics}. 
Then the non-Hermitian system is governed by Lindblad master equation \cite{1976.CMP},
\begin{equation}
\begin{aligned}
\frac{d \hat{\rho}}{d t}= 
-\frac{i}{\hbar}[\hat{H}, \hat{\rho}]
&+\sum_{j=1,2}\gamma_j(2 \hat{a}_j \hat{\rho} \hat{a}_j^{\dagger}
-\hat{\rho} \hat{a}_j^{\dagger} \hat{a}_j-\hat{a}_j^{\dagger} \hat{a}_j \hat{\rho})\\
&+\sum_{j=1,2}\beta_j(2 \hat{a}_j^{\dagger} \hat{\rho} \hat{a}_j
-\hat{\rho} \hat{a}_j \hat{a}_j^{\dagger}
-\hat{a}_j \hat{a}_j^{\dagger} \hat{\rho})
\end{aligned}
\end{equation}
where $\gamma_j$ ($\beta_j$) is the loss (gain) coefficient of the $j$th cavity. 
The effect of quantum jumping term  $\gamma_j(2 \hat{a_j} \hat{\rho} \hat{a}_j^{\dagger})$ coming from loss and $\beta_j(2 \hat{a}_j^{\dagger} \hat{\rho} \hat{a}_j)$ from gain on the quantum behavior is totally different \cite{2018.EPL}. 
While  in classical PT-symmetry or -broken systems \cite{2010.NatPhys}, these two effects are looked as an average one, where active materials  with a loss (such as the scattering and absorption) can be compensated by a gain.

\begin{figure*}[htb]
  \centering
  \includegraphics[width=\textwidth]{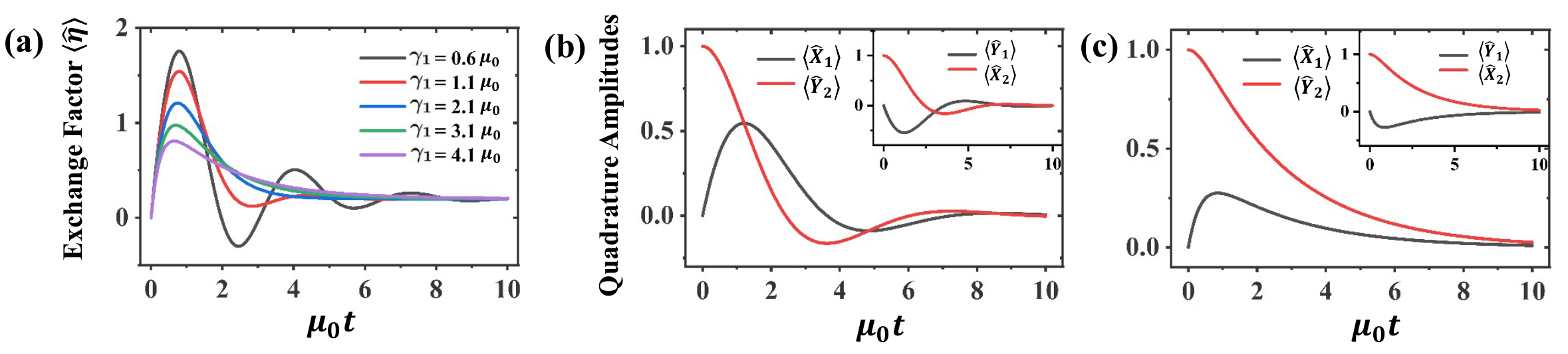}\\
  \caption{
 Characterizing the transmission from PT-symmetry to PT-broken phase through the exchange factor $\langle\hat{\eta}\rangle$. 
  (a) The evolution of $\langle\hat{\eta}\rangle$ with varying  $\gamma_1$ (normalized by $\mu_0$). 
The exchange between $\langle\hat{X}_{1}\rangle$ and $\langle\hat{Y}_{2}\rangle$ (inset: $\langle\hat{X}_{2}\rangle$ and $\langle\hat{Y}_{1}\rangle$ ) (b) in PT-symmetry with  $\gamma_1=1.1\mu_0$  (yellow star in Fig. 1(b)) and 
  (c) in PT-broken with $\gamma_1=3.1\mu_0$ (red star in Fig. 1(b)).
The initial state  is coherent state $|0,\alpha=1+i\rangle$ and other parameters are $ \beta_1=0.1\mu_0, \gamma_2=0.1\mu_0, \beta_2=0.1\mu_0$, and $\mu=\mu_0$. }
\label{fig2}
\end{figure*}

To construct the quantum PT-phase diagram, based on  Eq. (2), we derive the evolution  of $\langle\hat{a}_{1}\rangle$ and $\langle\hat{a}_{2}\rangle$ with varying $t$  \cite{SM}, 
\begin{equation}
\begin{gathered}
i\frac{d}{d t}\left(\begin{array}{l}
\langle\hat{a}_{1}\rangle \\
\langle\hat{a}_{2}\rangle
\end{array}\right)=H_{\mathrm{eff}}\left(\begin{array}{l}
\langle\hat{a}_{1}\rangle \\
\langle\hat{a}_{2}\rangle
\end{array}\right)
\end{gathered}
\end{equation}
where
\begin{equation}
\begin{gathered}
H_{\mathrm{eff}}=\left(\begin{array}{cc}
\omega-i\gamma_{1}+i\beta_{1} & \mu \\
\mu & \omega-i\gamma_{2}+i\beta_{2}
\end{array}\right).
\end{gathered}
\end{equation}
The eigenvalues of  $H_{\mathrm{eff}}$ are
\begin{equation}
\begin{aligned}
\omega_{\pm}&=\omega-\frac{i}{2}(\gamma_{1}-\beta_{1}+\gamma_{2}-\beta_{2})\\
&\pm\frac{1}{2}\sqrt{4\mu^{2}-[(\gamma_{1}-\beta_{1})-(\gamma_{2}-\beta_{2})]^{2}}.
\end{aligned}
\end{equation}
The degeneracy parts of eigenvalues $\omega_{\pm}$, which satisfy $(\gamma_{1}-\beta_{1})/\mu-(\gamma_{2}-\beta_{2})/\mu=\pm2$, are called EP lines, shown as two red lines in PT-phase diagram [Fig. 1(b)]. 
The area between two red lines  is PT-symmetric while the areas outside these two lines are PT-broken. 
PT-phase diagram in quantum system is different from that in the classical system \cite{2023.JOP}, where the effect of gain and loss is averaged by $(\gamma_j-\beta_j)$.
In the quantum phase diagram, each single point corresponds to countless options of $\gamma_j$ and $\beta_j$ but with a fixed value of  $(\gamma_j-\beta_j)$.

On the other hand, Eq. (2) can be written as $\frac{d \hat{\rho}}{d t}=\mathcal{L}\hat{\rho}$ with the  Liouvillian operator $\mathcal{L}$.
Given a set of complete quantum state basis vectors,
 $\mathcal{L}$ can be expressed as a high dimension matrix. 
Fig. 1(c) shows the real parts of eigenvalues of Liouvillian $\mathcal{L}(\mu)$ with $\gamma_1=3.1\mu_0, \beta_1=0.1\mu_0, \gamma_2=1.1\mu_0$, and $\beta_2=0.1\mu_0$, respectively.  
One can see that the splitting point locating at $\mu=\mu_0$ is identical to the gray star in the phase diagram [Fig. 1(b)].
More details about $\mathcal{L}$ are shown in Ref. \cite{SM}.

Furthermore, the evolution of the mean photon number $\langle\hat{n}_1\rangle=\langle\hat{a}_1^\dagger\hat{a}_1\rangle$, $\langle\hat{n}_2\rangle=\langle\hat{a}_2^\dagger\hat{a}_2\rangle$ of two modes, and an exchange factor $\langle\hat{\eta}\rangle=\langle i(\hat{a}_{2}^\dagger\hat{a}_{1}-\hat{a}_{1}^\dagger\hat{a}_{2})\rangle$ can be written as  \cite{SM},
\begin{equation}
\begin{aligned}
\frac{d}{d t} \langle \hat{n}_{1} \rangle =&\,2\left(\beta_{1}-\gamma_{1}\right) \langle \hat{n}_{1} \rangle +\mu \langle \hat{\eta} \rangle+2\beta_{1}\\
\frac{d}{d t} \langle \hat{n}_{2} \rangle =&\,2\left(\beta_{2}-\gamma_{2}\right) \langle \hat{n}_{2} \rangle -\mu \langle \hat{\eta} \rangle+2\beta_{2}\\
\frac{d}{d t}\langle \hat{\eta} \rangle =&\,2\mu\langle \hat{n}_{2} \rangle-2\mu\langle \hat{n}_{1}\rangle\\
& +\left(\beta_{1}+\beta_{2}-\gamma_{1}-\gamma_{2}\right) \langle \hat{\eta} \rangle
\end{aligned}
\end{equation}
whose solutions satisfy the steady state conditions that both $\gamma_{1}+\gamma_{2}-\beta_{1}-\beta_{2}>0$ and $(\gamma_{1}-\beta_{1})(\gamma_{2}-\beta_{2})+\mu^{2}>0$,  shown as the yellow area of phase diagram in Fig. 1(b). 
Under the steady state conditions, the final values of mean photon number of two modes as well as $\langle\hat{\eta}\rangle$ can be written as  \cite{SM},
\begin{equation}
\begin{aligned}
\left\langle \hat{n}_{1} \right\rangle_{s s}&=\frac{\Delta_{1}-\beta_{1}\left(\Delta_{2}+2\beta_{2}\gamma_{2}-\gamma_{2}^{2}\right)}{\Delta_{3}}\\
\left\langle \hat{n}_{2} \right\rangle_{s s}&=\frac{\Delta_{1}-\beta_{2}\left(\Delta_{2}+2\beta_{1}\gamma_{1}-\gamma_{1}^{2}\right)}{\Delta_{3}}\\
\left\langle \hat{\eta} \right\rangle_{s s}&=\frac{2\mu\left(\beta_{2}\gamma_{1}-\beta_{1}\gamma_{2}\right)}{\Delta_{3}}
\end{aligned}
\end{equation}
with $\Delta_1=(\beta_1+\beta_2)(\beta_1\beta_2+\mu^2)$, 
$\Delta_2=\beta_1\gamma_2+\beta_2\gamma_1-\gamma_1\gamma_2$,
and $\Delta_3=(\gamma_1+\gamma_2-\beta_1-\beta_2)[(\gamma_1-\beta_1)(\gamma_2-\beta_2)+\mu^2]$. 
Here, the parameters in the steady state region should be satisfied with the condition of weak gain.
From Eq. (7), one can see that, for one steady state point, there are infinite sets of parameters $\gamma_1$, $\beta_1$, $\gamma_2$, and $\beta_2$ corresponding to infinite steady state values. 
But owing to the decoherence effects of loss and gain, the steady state will finally become a thermal state without any quantum feature \cite{SM}. 
Our following discussions are limited within the steady state regime.


{\it{Exchange operator to characterize PT-phase.}}
To characterize the PT-symmetry or -broken, we rewrite the exchange operator $\hat{\eta}$ as
\begin{equation}
\hat{\eta}=2(\hat{X}_1\hat{Y}_2-\hat{X}_2\hat{Y}_1)
\end{equation}
with $\hat{X}_{1,2}=(\hat{a}_{1,2}+\hat{a}^\dagger_{1,2})/2$ and $\hat{Y}_{1,2}=(\hat{a}_{1,2}-\hat{a}^\dagger_{1,2})/2i$. 
$\hat{\eta}$, as an Hermitian operator, expresses the exchanging between quadrature  variables $\hat{X}_{1,2}$ and $\hat{Y}_{2,1}$.
 Its expectation value $\langle\hat{\eta}\rangle$ is a real number,  called exchange factor.
Fig. 2(a) gives the evolution of $\langle\hat{\eta}\rangle$ with varying the loss rate $\gamma_1$.
Here, $ \beta_1=0.1\mu_0, \gamma_2=0.1\mu_0, \beta_2=0.1\mu_0, \mu=\mu_0$, where $\mu_0=1\times10^{-10}$ Hz, and the initial state  is a coherent state $|0,\alpha=1+i\rangle$.
From Fig. 2(a), $\langle\hat{\eta}\rangle$  experiences the phase transition from PT-symmetry at $\gamma_1=1.1\mu_0$, via the EP point at $\gamma_1=2.1\mu_0$, to PT-broken at $\gamma_1=3.1\mu_0$, corresponding to the yellow, gray, and red stars in Fig. 1(b), respectively.
It is seen that, when PT-symmetry is unbroken, $\langle\hat{\eta}\rangle$  oscillates with  $t$. In contrast, when PT-symmetry is broken, $\langle\hat{\eta}\rangle$ monotonically decreases after a rise and then comes to the steady state.
It is noted that whenever for any value of $\gamma_1$, $\langle\hat{\eta}\rangle$ is approaching to the same value when $t\rightarrow\infty$. This is not a general case, but an accident just for the condition of $\gamma_2=\beta_2$.

\begin{figure*}[htb]
  \centering
  \includegraphics[width=\textwidth]{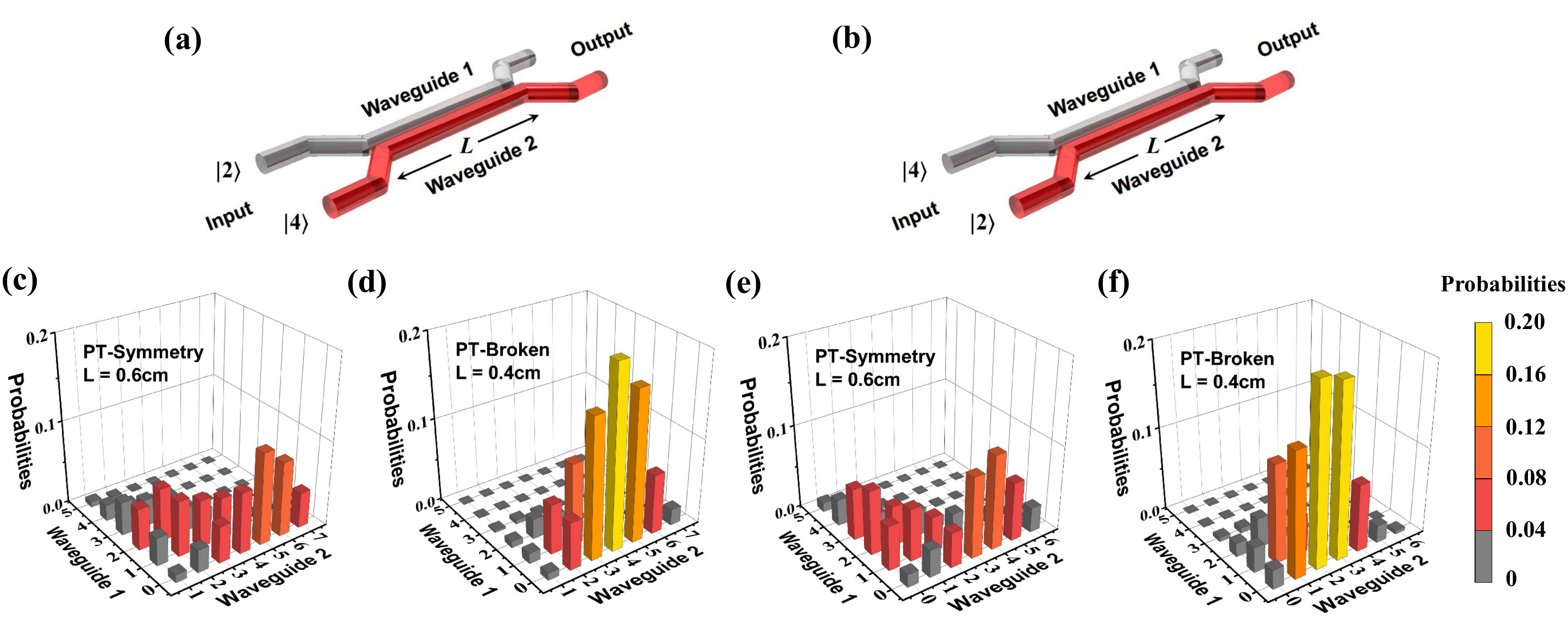}\\
 \caption{Quantum state engineering  based on PT-symmetry or PT-broken. 
 Schematics of coupled waveguides with input Fock states (a) $|2,4\rangle$ for (c, d) and (b) $|4,2\rangle$ for (e, f). 
Probability distributions of output states  for (c, e) PT-symmetry at $L=0.6$ cm and (d, f) PT-broken at $L=0.4$ cm.
Other parameters are the same as those in Figs. 2(b, c).}
\label{fig3}
\end{figure*}

Correspondingly, we explore the exchanging processing  between quadrature amplitudes $\langle\hat{X}_{1,2}\rangle$ and $\langle\hat{Y}_{2,1}\rangle$. 
For the PT-symmetry, there is a periodical exchanging process between $\langle\hat{X}_{1,2}\rangle$ and $\langle\hat{Y}_{2,1}\rangle$, as shown in Fig. 2(b) with $\gamma_1=1.1\mu_0$.
In contrast, when  PT is broken, they decay exponential with $\langle\hat{X}_{1}\rangle<\langle\hat{Y}_{2}\rangle$ and $\langle\hat{Y} _{1}\rangle<\langle\hat{X}_{2}\rangle$, as shown in Fig. 2(c) with $\gamma_1=3.1\mu_0$.
If now, we input the Fock state $\ket{m,n}$ as an initial state, $\langle\hat{X}_{1,2}\rangle$ and $\langle\hat{Y}_{2,1}\rangle$ will be $0$ for all the time \cite{SM}.
The reason is that the average values of $\langle\hat{X}_{1,2}\rangle$ and $\langle\hat{Y}_{2,1}\rangle$ in the Fock states are always zero.
So if only inputting Fock states,  one can not use the exchange of quadrature amplitudes to distinguish  the PT-symmetry or -broken. 
While, whatever for Fock states or coherent states, one can clearly distinguish them through the exchange factor $\langle\hat\eta\rangle$ \cite{SM}.
Therefore, exchange operator $\hat\eta$ can fully characterize the properties of PT-symmetry or broken in quantum photonic system.

{\it{Engineering quantum state with PT-broken.}}
The above theory can be applied to any two-mode coupling photonic structures. 
If existing the  loss and gain  in the photonic structure, it can be equivalent to a non-Hermitian beam splitter \cite{1998.PRA}.
Non-Hermitian beam splitters have some unique properties and applications, such as quantum coherent perfect absorption \cite{2016.PRL}, anti-bunching of bosons \cite{2017.Science}, preparation of squeezed states \cite{2019.SPIE}, and fabrication of multi-bit quantum gates \cite{2021.JOSAB}. 
Now, let's take coupled waveguide system as an example to study the quantum state engineering. 
Shown as Figs. 3(a, b), two gain-loss waveguides with coupled distance $L$ can be looked as a non-Hermitian beam splitter.
If $L$ is too short, the interaction between two modes is not enough. 
In contrast, if $L$ is too long, any input quantum state will become a thermal state. 
Thus there exists an optimal interval of $L$, in which the interaction between two modes is enough while  quantum coherence and PT-symmetry come into the effect together.
When $\gamma_1=1.1\mu_0\sim3.1\mu_0, \beta_1=0.1\mu_0, \gamma_2=0.1\mu_0, \beta_2=0.1\mu_0$, and $\mu=\mu_0$, where $\mu_0=1\,\mathrm{cm}^{-1}$\cite{2010.NatPhys,2019.NP}, the optimal value of $L$ is $0.4\sim1.5$ cm.
With the above parameters, in the following discussions, we will focus on the quantum state distribution of two outputs  for both PT-symmetry  and PT-broken cases.

\begin{table}
\caption{Probability distributions and mean photon numbers of output state in the non-Hermitian  beam splitters.}
\label{tab:1}
\centering
\subtable[\,PT-symmetry]{
\begin{ruledtabular}
\begin{tabular}{c|c|ccccccccc}
 Input & \textit{L}/cm&$\ket{0,1}$ & $\ket{1,0}$&$\ket{0,2}$&$\ket{1,1}$
 &$\ket{2,0}$ &$\ket{0,3}$&...&$\langle\hat{n}_1\rangle$ &$\langle\hat{n}_2\rangle$\\
\hline
$\ket{0,1}$ & 1.2 & 0.21 & 0.21 & 0.06 & 0.06 & 0.04 & 0.01 &...& 0.41 & 0.51\\
$\ket{1,0}$ & 1.2 & 0.25 & 0.06 & 0.07 & 0.02 & 0.01 & 0.02 &...& 0.11 & 0.51\\
$\ket{0,2}$ & 0.75 & 0.2 & 0.08 & 0.23 & 0.17 & 0.04 & 0.08 &...& 0.54 & 1.4\\
$\ket{1,1}$ & 0.75 & 0.24 & 0.18 & 0.19 & 0.04 & 0.05 & 0.06 &...& 0.41 & 1.0\\
$\ket{2,0}$ & 0.75 & 0.25 & 0.13 & 0.08 & 0.05 & 0.02 & 0.02 &...& 0.28 & 0.6\\
\end{tabular}
\end{ruledtabular}
}
\subtable[\,PT-broken]{
\begin{ruledtabular}
\begin{tabular}{c|c|ccccccccc}
 Input & \textit{L}/cm&$\ket{0,1}$ & $\ket{1,0}$&$\ket{0,2}$&$\ket{1,1}$
 &$\ket{2,0}$ &$\ket{0,3}$&...&$\langle\hat{n}_1\rangle$ &$\langle\hat{n}_2\rangle$\\
\hline
$\ket{0,1}$ & 0.7 & 0.52 & 0.06 & 0.12 & 0.03 & 0 & 0.02 &...& 0.11 & 0.89\\
$\ket{1,0}$ & 0.7 & 0.14 & 0.03 & 0.02 & 0.01 & 0 & 0 &...& 0.04 & 0.21\\
$\ket{0,2}$ & 0.5 & 0.21 & 0.02 & 0.47 & 0.08 & 0 & 0.12 &...& 0.16 & 1.79\\
$\ket{1,1}$ & 0.5 & 0.53 & 0.07 & 0.16 & 0.03 & 0.01 & 0.03 &...& 0.13 & 1.01\\
$\ket{2,0}$ & 0.5 & 0.15 & 0.07 & 0.02 & 0.01 & 0 & 0 &...& 0.09 & 0.22\\
\end{tabular}
\end{ruledtabular}
}
\end{table}

We first consider the situation of single photon input, i.e., $|\psi\rangle_{in}=|0,1\rangle$ and $|1,0\rangle$ \cite{SM}. 
In the case of PT-symmetry, corresponding to the yellow star in Fig. 1(b), at $L=1.2$ cm, probability distributions and mean photon numbers of output states are shown in Tab. I(a). 
It is seen that the photons tend to symmetrically distribute in the two waveguides. 
While in the PT-broken case, the red star in Fig. 1(b), output states at $L=0.7$ cm are  shown in Tab. I(b).  
In this case, whatever inputting one photon from which waveguide, the photons are likely to output from the dominant waveguide.
The result of PT-broken  is in agreement with the classical optical experiments where most of the energy is locating in the dominant mode \cite{2010.NatPhys}.

Then the situation of two-photon input is explored,  i.e., $|\psi\rangle_{in}=|0,2\rangle$, $|1,1\rangle$, and $|2,0\rangle$  \cite{SM}.
For the PT-symmetry, two output states at $L=0.75$ cm are shown in Tab. I(a), while  for the PT-broken, two output states at $L=0.5$ cm are shown in Tab. I(b).
For both cases, the probabilities of output state $|1,1\rangle$ are very small, shown as a dip in the probability distributions of photons with the distance $L$ \cite{SM}, i.e., the photons are  inclined to together output from one of the waveguides, appearing the results of HOM  \cite{1987.PRL, 2019.NP}.
Once again, from probability distributions and mean photon numbers of output states with both PT-symmetry and PT-broken,  the quantum results  are in accord with the corresponding classical ones  \cite{2010.NatPhys}.

The above two examples imply that the  beam splitter with the PT-symmetry is different from previous studied non-Hermitian one.
After EP,  most of photons (or the output states with large probability)  are concentrating on the dominant waveguide due to the joint effect of quantum interference  and PT-broken. 
Also, because of the existence of gain, the quantum state $\ket{P}$ with $P>(M+N)$ appears  when   $|\psi\rangle_{in}=\ket{M,N}$.
So, the beam splitter with this kind of PT-broken can be used to prepare the high number Fock state.
Now, we take the input states of $|\psi\rangle_{in}=|2,4\rangle$ and $|4,2\rangle$ as  examples.
As shown in Figs. 3(c, e), in the case of PT-symmetry (yellow star in Fig. 1(b)), the photon number distribution at $L=0.6$ cm is dispersed due to periodically exchanging between two waveguides. The mean photon numbers are $\langle \hat{n}_{1} \rangle=1.2$ and $\langle \hat{n}_{2} \rangle=3.3$ (input $\ket{2,4}$),  $\langle \hat{n}_{1} \rangle=1.1$ and $\langle \hat{n}_{2} \rangle=2.3$ (input $\ket{4,2}$). 
While in the PT broken (red star in Fig. 1(b)), at $L=0.4$ cm, most photons are gathered in the dominant waveguide with large probability distributions of high number Fock state  [Figs. 3(d, f)], corresponding to $\langle \hat{n}_{1} \rangle=0.4$ and $\langle \hat{n}_{2} \rangle=2.0$ (input $\ket{2,4}$), $\langle \hat{n}_{1} \rangle=0.3$ and $\langle \hat{n}_{2} \rangle=3.7$ (input $\ket{4,2}$).
We have checked other cases that input Fock state is $\ket{M,N}$ with the total number of photons $M+N<10$ and  the same conclusion is obtained \cite{SM}.
Moreover, by adjusting the loss and gain parameters of two waveguides, more optimized results about output states will appear.

{\it{Summary.}} We have analytically obtained the quantum PT-phase diagram with the steady state regime in non-Hermitian  photonic structures. 
We have defined an exchange operator to characterize the PT-symmetry phase and PT-broken phase.
Based on this phase diagram, we have engineered the multi-photon quantum state in the coupled waveguide structure.
The present work has constructed the basic theory of quantum PT-symmetry in photonic structure as well as its application to quantum state engineering.
The established theory can be extended to study many related quantum behaviors, such as gain saturation effect, quantum entanglement, and continuous variable states, and may have potential applications in quantum state preparation, quantum interferences, and  logic operations in non-Hermitian photonic systems.

\acknowledgments
\textit{Acknowledgments.} This work is supported by the National Natural Science Foundation of China under Grants Nos. 11974032 and by the Key R$\&$D Program of Guangdong Province under Grant No. 2018B030329001.

\end{document}